\documentclass[prl,twocolumn,nofootinbib,reprint,preprintnumbers,showkeys]{revtex4-1}
\usepackage{color}
\usepackage{amsmath,amsfonts,bm}
\usepackage{graphicx, epstopdf,scrextend}
\usepackage{mathrsfs,mathtools}

\definecolor{darkgreen}{rgb}{0,0.4,0} 
\definecolor{darkblue}{rgb}{0,0,0.6} 
\usepackage[colorlinks=true,linkcolor=darkblue,citecolor=darkgreen]{hyperref}

\newcommand{\LM}{\mathrm{M}}
\newcommand{\LL}{\mathrm{L}}

\newcommand{\LT}{\mathrm{T}}

\newcommand{\cL}{\mathcal{L}}

\newcommand{\cO}{\mathcal{O}}

\definecolor{red}{rgb}{1,0,0}

\def\<>#1{\big\langle{#1}\big\rangle}
\def\[]#1{\big[{#1}\big]}

\newif\ifusefigs
\usefigstrue

\begin{document}

\title{Sound waves move matter}

\author{Davison E. Soper}

\affiliation{
Institute of Theoretical Science,
University of Oregon,
Eugene, OR  97403-5203, USA
}

\email{soper@uoregon.edu}

\begin{abstract}
A recent paper has reported that sound waves can carry gravitational mass. I analyze this effect in a Hooke's law solid, considering a wave packet moving in the $z$ direction with an amplitude that is independent of $x$ and $y$. The analysis shows that, at second order in an expansion around small amplitude vibrations, there is a small net motion of material, and thus mass, in the direction opposite to the wave packet propagation. This is a straightforward consequence of Newton's laws.
\end{abstract}

\date{4 November 2019}

\maketitle
\section{Introduction}

In a recent paper \cite{EKN}, Esposito, Krichevsky, and Nicolis have argued that sound waves can carry gravitational mass. In the simplest approximation, a sound wave obeys a linear wave equation and carries energy. If a sound wave packet carries energy $E$, then it evidently carries mass $M_0 = E/c^2$. However, this tiny mass is not the mass considered. Rather, the equation of motion for a real material is actually nonlinear. Considering the first nonlinear corrections to the motion of the material, there is a much larger mass associated with the wave packet, $M \propto E/c_\LL^2$, where $c_\LL$ is the speed of sound in the material \cite{EKN}. This result generalizes an analogous result \cite{Nicolis:2017eqo} for phonons in a superfluid.

In this paper, I analyze this effect. I consider sound propagation in an isotropic Hooke's Law solid, using the classical field theory for continuum mechanics as formulated in Ref.~\cite{CFT}. I use the non-relativistic formulation of the theory because, in my opinion, the description is simpler. I consider a sound wave packet that moves in the $+z$ direction and is infinite in extent in the $x$ and $y$ directions. The oscillations are in the $\pm z$ direction and have an amplitude that depends on $t$ and $z$ but is independent of $x$ and $y$. Instead of having a solid that is infinite in extent in $x$ and $y$, one could equally well consider a soft solid that is contained in a rigid tube oriented in the $z$ direction with sliding boundary conditions. 

The analysis requires a detailed examination of the equations of motion. However, the physical effect is very simple and can be understood without any equations at all. A sound wave propagates because the atoms in the material are subject to forces. A given atom is at rest before the wave packet reaches it. As the wave packet is passing, the atom is subject to forces, which cause it to move. After the wave packet has passed, the solution of the equations of motion shows, not surprisingly, that no net momentum has been transferred to the atom. However, the fact that the atom is again at rest does not necessarily imply that it winds up where it started. In fact, considering the nonlinear terms in the equations of motion, one finds that the atom has moved slightly in the $-z$ direction. This is a very small effect, but it is not zero. 

There is another way to look at this. Consider two atoms in the material, one at $z = z_1$ and one at $z = z_2 > z_1$. The wave packet arrives at atom 1 first and (averaged over vibrations) it starts moving backward, in the $- z$ direction. Then the wave packet arrives at atom 2. It also starts moving backward. But atom 1 has already moved some distance, so the distance between the two atoms has increased. That is, the number of atoms per unit volume has decreased. Thus the mass density inside the wave packet is smaller than the undeformed mass density. We can characterize this by saying that a mass deficit or a negative mass density is carried along with velocity $c_\LL$ by the wave packet. This is the interpretation given in Ref.~\cite{EKN}. 

\section{The Lagrangian}

After this introduction, we are ready to analyze the sound wave. We follow quite closely the notation of Ref.~\cite{CFT} in which continuum mechanics is considered as a classical field theory based on a Lagrangian. Expositions of parts of this formulation of continuum mechanics can be found in Refs.~\cite{EKN,Dubovsky:2005xd,Endlich:2010hf,Endlich:2012pz}.

Although we will use a non-relativistic Lagrangian, it is convenient to use a notation with space-time coordinates $x = (x^0, x^1, x^2, x^3) = (t,x,y,z)$. We use a metric $g^{\mu\nu} = \textrm{diag}(-1,+1,+1,+1)$ to raise and lower indices.\footnote{For ``0'' indices, I mostly use upper indices except for $\partial_0 = \partial/\partial x^0$. For ``1,2,3'' indices, upper and lower indices are equivalent.} Repeated Latin indices are summed from 1 to 3, while repeated Greek indices are summed from 0 to 3. To begin, consider labels $R = (R_1, R_2, R_3)$ carried by the atoms in the material. We consider a continuum description of the material, so that discrete atoms do not really play a role in the  equations except as a device to ascribe labels fixed to the material. Call the labels that appear at a given space time point $R_a(x)$. The functions $R_a(x)$ are then our fields.

We start with a solid in an unstressed, lowest energy state. For this state, we assign the labels $R_a$ so that $R_a(x) = x^a$ for $a = 1,2,3$.

Following the notation of Ref.~\cite{CFT}, we let $n$ be the number of atoms per unit $d^3R$ and $m$ be the mass per atom, so that $mn$ is the mass per unit $d^3R$. We define the conserved matter current $J^\mu$ by  Eq.~(4.1.2) of Ref.~\cite{CFT}
\begin{equation}
J^\mu = \frac{n}{3!}\epsilon^{\mu\alpha\beta\gamma}\epsilon_{abc}
(\partial_\alpha R_a)(\partial_\beta R_b)(\partial_\gamma R_c)
\;.
\end{equation}
Thus $J^0$ is $n$ times the determinant of $\partial_\alpha R_a$, so that $J^0$ is the number of atoms per unit $d^3 \vec x$ and $J^i$ is the number of atoms crossing a surface oriented perpendicular to the $i$ direction per unit area per unit time. The velocity of the material is then
\begin{equation}
v^i = \frac{J^i}{J^0}
\;.
\end{equation}
We define a (non-relativistic) deformation matrix by
\begin{equation}
\widetilde G_{ab} = (\partial_k R_a)(\partial_k R_b)
\;.
\end{equation}
(The relativistic version of this, with $c = 1$, is $G_{ab} = (\partial_\mu R_a)(\partial^\mu R_b)$, but we drop the time derivatives of the fields because they are tiny in terrestrial environments.) When the material is undeformed, so that $R_a = x_a$, $\widetilde G_{ab}$ is the unit matrix. For a deformed material, we define the strain matrix from Eq.~(6.1.2) of Ref.~\cite{CFT},
\begin{equation}
s_{ab} = \frac{1}{2}\left(\delta_{ab} - \widetilde G_{ab}\right)
\;.
\end{equation}

The non-relativistic Lagrangian from Eq.~(7.1.8) of Ref.~\cite{CFT} is
\begin{equation}
\cL = \frac{1}{2}\, m J^0 \bm v^2
- J^0 \widetilde U(\widetilde G_{ab},R_a)
\;. 
\end{equation}
Here $(1/2) m J^0 \bm v^2$ is the density of kinetic energy of the material and $\widetilde U(\widetilde G_{ab},R_a)$ is the potential energy per unit $d^3 R$ of the material, so that $J^0 \widetilde U$ is the potential energy per unit $d^3 x$.

Here $\widetilde U(\widetilde G_{ab},R_a)$ could be any function of $\widetilde G_{ab}$ and $R_a$. We will consider a uniform, isotropic, Hooke's Law material, for which, from Sec.~6.3 of Ref.~\cite{CFT},
\begin{equation}
\label{eq:HookesLaw}
\widetilde U = \frac{1}{2n}\left[\lambda (s_{aa})^2 + 2 \mu\, s_{ab}s_{ab}\right]
\;.
\end{equation}
Here $\lambda$ and $\mu$ are two elastic constants, the Lam\'e constants.

Since the action is invariant under time translations, energy is conserved in the absence of external forces: $\partial_0 T^{00} + \partial_k T^{0k} = 0$. The energy density is (from Eq.~(7.2.5) of Ref.~\cite{CFT})
\begin{equation}
\label{eq:T00}
T^{00} = \frac{1}{2} m J^0 \bm v^2
+ J^0 \widetilde U
\;.
\end{equation}
The energy current (from Eq.~(7.2.6) of Ref.~\cite{CFT}) is
\begin{equation}
\begin{split}
T^{0k} ={}&  \left(\frac{1}{2} m J^0 \bm v^2
+ J^0 \widetilde U \right)v^k
\\ & + 2 J^0 (\partial_j R_a)(\partial_k R_b)\frac{\partial \widetilde U}{\partial G_{ab}}\
v^j
\;.
\end{split}
\end{equation}

Since the action is invariant under space translations, momentum is conserved if there are no external forces. If there is an applied force density $f$, then 
\begin{equation}
\label{eq:momentumconservation}
\partial_0 T^{j0} + \partial_k T^{jk} = f^j
\;.
\end{equation}
We will use this momentum conservation equation as the equation of motion for the material. The momentum density is (from Eq.~(7.2.3) of Ref.~\cite{CFT})\footnote{Note that $T^{j0} = T^{0j}$ in relativistic mechanics as a consequence of Lorentz invariance, but that $T^{j0} \ne T^{0j}$ in the non-relativistic case.}
\begin{equation}
\label{eq:momentumdensity}
T^{j0} = m J^0 v^j = m J^j
\;.
\end{equation}
The momentum current, that is, the stress tensor, is (from Eq.~(7.2.4) of Ref.~\cite{CFT})
\begin{equation}
\begin{split}
\label{eq:stresstensor}
T^{jk} ={}&  m J^0 v^j v^k
+2 J^0 (\partial_j R_a)(\partial_k R_b)\frac{\partial \widetilde U}{\partial G_{ab}}
\;.
\end{split}
\end{equation}
For our isotropic Hooke's law material, this is
\begin{equation}
\begin{split}
\label{eq:Tjk}
T^{jk} ={}&  m J^0 v^j v^k
\\&
- \frac{1}{n} J^0 (\partial_j R_a)(\partial_k R_b)
[\lambda s_{cc}\delta_{ab} + 2 \mu s_{ab}]
\;.
\end{split}
\end{equation}
%

\section{Equations of motion for sound waves}

To analyze sound waves, we let
\begin{equation}
R_a = x_a + \varphi_a
\;,
\end{equation}
where $\varphi(x)$ is small. The equation of motion for $\varphi(x)$ is non-linear. To proceed, we use perturbation theory. The sound wave in $\varphi(x)$ is created by a force density $f(x)$ according to the equation of motion (\ref{eq:momentumconservation}). We expand $\varphi(x)$ in powers of $f$. We will let
\begin{equation}
\label{eq:phiexpansion}
\varphi_a(x) = \phi_a(x) + \psi_a(x) + \cO(f^3)
\;,
\end{equation}
where $\phi_a(x)$ is proportional to $f$ and $\psi_a(x)$ is proportional to $f^2$.

One might try to analyze a wave form $\varphi(x)$ of arbitrary shape in three dimensions, as in Ref.~\cite{EKN}. However we will specialize to a case with much higher symmetry. We consider a field $\varphi(x)$ and a force $f(x)$ that are invariant under translations in the $x$ and $y$ directions and under rotations about the $z$ axis. Then each of $f$ and $\varphi$ have only one non-zero component, $f_3$ and $\varphi_3$, and this component is a function only of $t$ and $z$.

With this simple geometry, we can find the behavior of $\varphi_3(x)$ by using conservation of the 3-component of momentum
\begin{equation}
\label{eq:momentumconservation3}
\partial_0 T^{30} + \partial_3 T^{33} = f_3
\;.
\end{equation}
From Eq.~(\ref{eq:momentumdensity}), the momentum density is
\begin{equation}
T^{30} = - mn \partial_0 \phi_3 - mn \partial_0 \psi_3 + \cO(f^3)
\;.
\end{equation}
From Eq.~(\ref{eq:Tjk}), the stress $T^{33}$ is
\begin{equation}
\begin{split}
\label{eq:T33}
T^{33} ={}& m n c_\LL^2 (\partial_3 \phi_3) 
+ m n c_\LL^2 (\partial_3 \psi_3) 
\\& + mn (\partial_0 \phi_3)^2
+ \frac{7}{2}\, m n c_\LL^2
(\partial_3 \phi_3)^2
+ \cO(f^3)
\;.
\end{split}
\end{equation}
Here
\begin{equation}
\begin{split}
\label{eq:cL}
c_\LL ={}& \left[\frac{\lambda + 2 \mu}{mn}\right]^{1/2}
\end{split}
\end{equation}
is the speed of longitudinal waves in the material. The speed of transverse waves is 
\begin{equation}
\label{eq:cT}
c_\LT = \left[\frac{\mu}{mn}\right]^{1/2}
\;.
\end{equation}
%


The equation of motion (\ref{eq:momentumconservation3}) gives us
\begin{equation}
\left[
\frac{1}{c_\LL^2}\partial_0 \partial_0 - \partial_3 \partial_3
\right]
\phi_3
= - \frac{f_3}{mn c_\LL^2}
\end{equation}
and
\begin{equation}
\label{eq:psi3PDE}
\left[
\frac{1}{c_\LL^2}\partial_0 \partial_0 - \partial_3 \partial_3
\right]
\psi_3
= 
\partial_3
\left[
\frac{1}{c_\LL^2}(\partial_0 \phi_3)^2
+\frac{7}{2} (\partial_3 \phi_3)^2
\right]
\;.
\end{equation}
%

\section{Solution of the equations of motion}

To aid in solving the equations of motion, define coordinates
\begin{equation}
x^\pm = c_\LL t \pm z
\;.
\end{equation}
Then
\begin{equation}
\begin{split}
c_\LL t ={}& \frac{1}{2}\left[x^+ + x^- \right]
\;,
\\
z ={}& \frac{1}{2}\left[x^+ - x^- \right]
\;.
\end{split}
\end{equation}
Also
\begin{equation}
\begin{split}
\frac{1}{c_\LL}\, \partial_0 ={}& \partial_+ + \partial_-
\;,
\\
\partial_3 ={}& \partial_+ - \partial_-
\;.
\end{split}
\end{equation}
Now our equations of motion are
\begin{equation}
\label{eq:phi3equation}
\partial_+\partial_- \phi_3 
= - \frac{1}{4 mn c_\LL^2}\ f_3
\end{equation}
and
\begin{equation}
\label{eq:psi3equation}
\partial_+\partial_- \psi_3 
= (\partial_+ - \partial_-)\, A
\;,
\end{equation}
where we have defined
\begin{equation}
\label{eq:Adef}
A = \frac{1}{4c_\LL^2}\,(\partial_0 \phi_3)^2
+\frac{7}{8}\, (\partial_3 \phi_3)^2
\;.
\end{equation}

We assume that $f_3(t,z) \ne 0$ only for $-R < x^+ < R$ and $-R < x^- < R$ for some distance parameter $R$. We impose the additional condition 
\begin{equation}
\label{eq:fcondition}
\int_{-\infty}^{\infty}\!d \bar x^- \ f_3(x^+, \bar x^-) = 0
\;.
\end{equation}
For instance, this condition holds if $f_3(x^+, - x^-) = -f_3(x^+, x^-)$. For a definite example, one could choose
\begin{equation}
\begin{split}
\label{eq:f3example}
f_3(x^+, x^-)
={}& \frac{4 m n c_\LL^2 f_0}{2R}\, 
\theta(-R < x^+ < R)
\\&\times
\theta(-R < x^- < R)\, \sin(k x^-)
\;,
\end{split}
\end{equation}
where $f_0$ is a small dimensionless parameter. The wave number $k$ is arbitrary, but $k = N\pi/R$ for a large integer $N$ would be a sensible choice. This is only an example. Any function that vanishes outside of $-R < x^\pm < R$ and obeys Eq.~(\ref{eq:fcondition}) will do.

We will assume that $\varphi(x)$ vanishes for times before the source $f(x)$ was on. Then $\varphi(x)$ must vanish outside of the forward ``soundcone'' from the region in which $f(x)$ was on: $\phi(x^+,x^-) = \psi(x^+,x^-) = 0$ for $x^+ < -R$ and for $x^- < -R$.

We can solve the equation of motion (\ref{eq:phi3equation}) for $\phi_3$ if we use the fact that $\phi_3$ vanishes for $x^\pm < - R$:
\begin{equation}
\phi_3(x^+,x^-) = 
- \frac{1}{4mn c_\LL^2}
\int_{-\infty}^{x^+}\!d \bar x^+ 
\int_{-\infty}^{x^-}\!d \bar x^- \ f(\bar x^+, \bar x^-)
\;.
\end{equation}

Consider the region $x^- > R$. In this region, we can replace the upper limit of the $\bar x^-$ integration by infinity, giving
\begin{equation}
\begin{split}
\phi_3(x^+,x^-) ={}& 
- \frac{1}{4mn c_\LL^2}
\int_{-\infty}^{x^+}\!d \bar x^+ 
\int_{-\infty}^{\infty}\!d \bar x^- \ f(\bar x^+, \bar x^-)
\;.
\end{split}
\end{equation}
The condition (\ref{eq:fcondition}) ensures that the right hand side of this equation vanishes, so that
\begin{equation}
\phi_3(x^+,x^-) = 0, \hskip 1 cm x^- > R
\;.
\end{equation}

Now consider the region $x^+ > R$. In this region, we can replace the upper limit of the $\bar x^+$ integration by infinity, giving
\begin{equation}
\phi_3(x^+,x^-) = 
- \frac{1}{4mn c_\LL^2}
\int_{-\infty}^{\infty}\!d \bar x^+ 
\int_{-\infty}^{x^-}\!d \bar x^- \ f(\bar x^+, \bar x^-)
\;.
\end{equation}
In this region, then, we have a right moving sound wave, that is, a function of $x^- = c_L t - z$ alone. 

\begin{figure}
\begin{center}
\includegraphics[width = 6 cm]{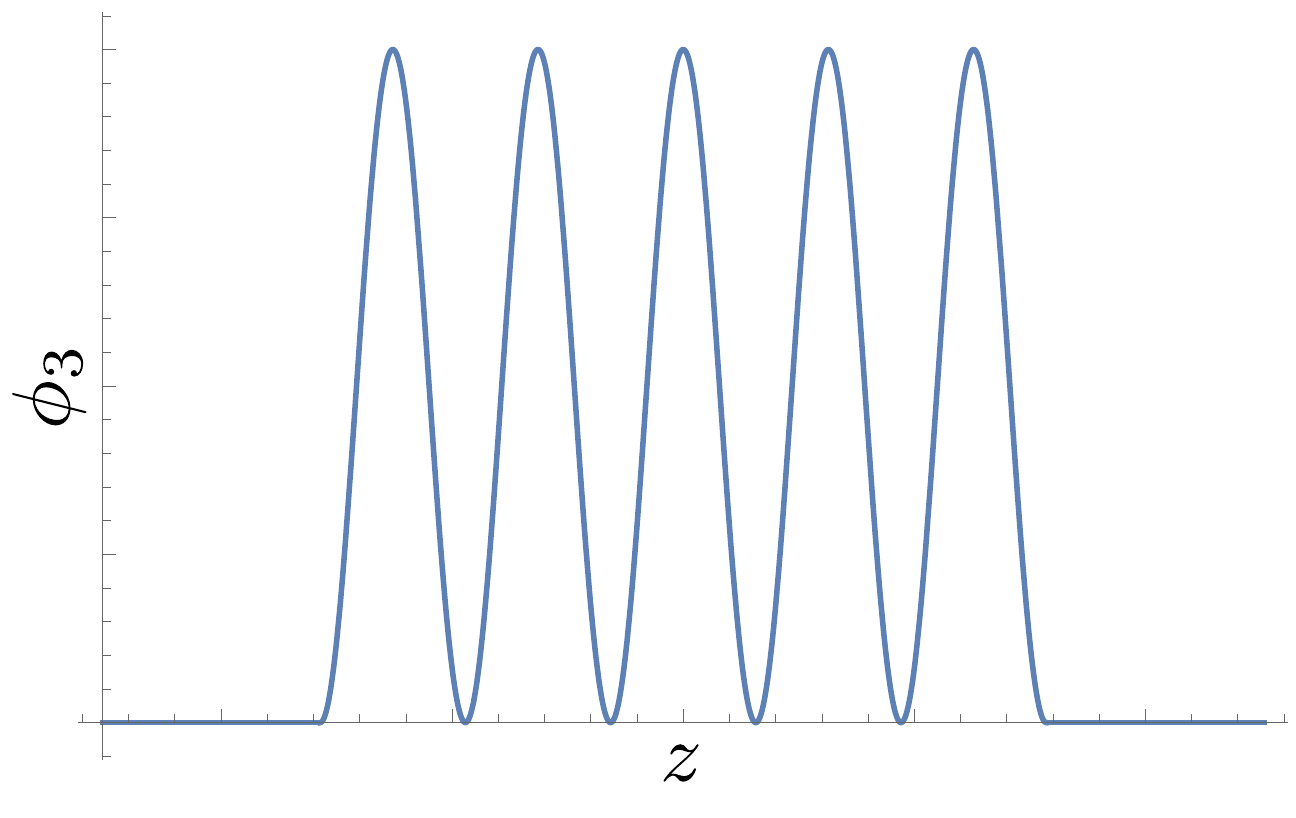}
\end{center}
\vskip -0.5 cm
\caption{
A sample wave packet.}
\label{fig:sample}
\end{figure}

For the example (\ref{eq:f3example}), $\phi_3$ in the region $x^+ > R$ is
\begin{equation}
\begin{split}
\label{eq:phi3example}
\phi_3(x^+, x^-)
={}& \frac{f_0}{k}\,\theta(-R < x^- < R)
\\&\times
[\cos(k x^-) - \cos(kR)]
\end{split}
\end{equation}
If we choose, say, $kR = 5 \pi$ then this is a wave packet with 5 oscillations, as shown in Fig.~\ref{fig:sample}. 

Next, we solve the equation of motion (\ref{eq:psi3equation}) for $\psi_3$, using the fact that $\psi_3$ vanishes for $x^\pm < - R$:
\begin{equation}
\begin{split}
\psi_3(x^+,x^-) ={}&
\int_{-\infty}^{x^+}\!d \bar x^+ 
\int_{-\infty}^{x^-}\!d \bar x^- \ 
\left(\frac{\partial}{\partial \bar x^+}
-\frac{\partial}{\partial \bar x^-}\right)
\\
&\times
A(\bar x^+, \bar x^-)
\;.
\end{split}
\end{equation}
Since $\phi_3(x^+, x^-)$, and thus $A(x^+, x^-)$ vanishes for $x^\pm < -R$, we can use integration by parts to write this as
\begin{equation}
\begin{split}
\psi_3(x^+,x^-) ={}&
\int_{-\infty}^{x^-}\!d \bar x^- \ 
A(x^+,\bar x^-)
\\
&
- \int_{-\infty}^{x^+}\!d \bar x^+ 
A(\bar x^+, x^-)
\;.
\end{split}
\end{equation}

Recall that $\phi_3(x^+,x^-)$ is independent of $x^+$ for $x^+ > R$. Thus
\begin{equation}
A(x^+,x^-) = A(R,x^-)
\hskip 1 cm x^+ > R
\;.
\end{equation}
We will examine $\psi_3(x^+,x^-)$ in the region $x^+ > R$. In this region, then, $\psi_3$ takes the form
\begin{equation}
\begin{split}
\label{eq:psi3result1}
\psi_3(x^+,x^-) ={}&
\int_{-\infty}^{x^-}\!d \bar x^- \ 
A(R,\bar x^-)
\\
&
-\int_{-\infty}^{R}\!d \bar x^+ 
A(\bar x^+, x^-)
\\&
- (x^+ - R)\,
A(R , x^-)
\;.
\end{split}
\end{equation}
We see that the second order wave $\psi_3(x^+,x^-)$ consists of three terms. The first and second terms describe right moving waves, that is, functions of $x^- = c_\LL t - z$ only. The third term describes a wave that is almost a purely right moving wave except that its amplitude is growing linearly with time. That is, it is a function of $x^-$ multiplied by a factor of $x^+ - R$. The primary wave is creating a disturbance, which is creating a right moving wave that moves along with the primary wave. Thus the amplitude of this wave increases with time. For sufficiently large values of $x^+$, $\psi_3$ will be as large as $\phi_3$ and our perturbative expansion will be spoiled, but we consider only the region of $x^+$ in which we can still use perturbation theory.

Eq.~(\ref{eq:psi3result1}) gives the solution $\psi_3(x^+,x^-)$ for $x^+ > R$. Now specialize further to $x^+ > R$ and $x^- > R$. Since $A(x^+,x^-) = 0$ for $x^- > R$, 
the last two terms in Eq.~(\ref{eq:psi3result1}) vanish and in the first term we can replace the upper endpoint of the $\bar x^-$ integration by infinity, giving
\begin{equation}
\begin{split}
\psi_3(x^+,x^-) ={}&
\int_{-\infty}^{\infty}\!d \bar x^- \ 
A(R,\bar x^-)
\;.
\end{split}
\end{equation}
This is a constant, independent of $t$ and $z$. The $x^+$ argument of $A(x^+,x^-)$ is indicated as $R$, but $A(x^+,x^-)$ is independent of $x^+$ in the region $x^+ > R$, so we can change the notation slightly to $A(\bar x^+,\bar x^-) = A(\bar x^-)$ and write this result as
\begin{equation}
\begin{split}
\psi_3 ={}&
\int_{-\infty}^{\infty}\!d \bar x^- \ 
A(\bar x^-)
\;.
\end{split}
\end{equation}
We choose a value of $t$ and let $\bar x^- = c_\LL t - z$, then change integration variable from $\bar x^-$ to $z$, giving
\begin{equation}
\begin{split}
\psi_3 ={}&
\int_{-\infty}^{\infty}\!d z \ 
A(c_\LL t - z)
\;.
\end{split}
\end{equation}
Using the definition (\ref{eq:Adef}) of $A$, this is
\begin{equation}
\begin{split}
\psi_3 ={}&
\int_{-\infty}^{\infty}\!d z 
\bigg[ 
\frac{1}{4c_\LL^2}\,(\partial_0 \phi_3(c_\LL t - z))^2
\\&\qquad
+\frac{7}{8}\, (\partial_3 \phi_3(c_\LL t - z))^2
\bigg]
\;.
\end{split}
\end{equation}
Since $\phi_3$ depends only on $c_\LL t - z$, this is
\begin{equation}
\begin{split}
\psi_3 ={}&
\frac{9}{8}\, \int_{-\infty}^{\infty}\!d x^3 
 (\partial_3 \phi_3)^2
\;.
\end{split}
\end{equation}
Note that this result does not depend on the particular example of a wave packet generated according to Eq.~(\ref{eq:f3example}).

We can compare this to the energy density of the primary wave. Eq.~(\ref{eq:T00}) gives
\begin{equation}
T^{00} = \frac{1}{2}\,mn c_\LL^2
\left[
\frac{1}{c_\LL^2}(\partial_0  \phi_3)^2
+(\partial_3  \phi_3)^2
\right]
\;,
\end{equation}
or
\begin{equation}
\label{eq:T00planewave}
T^{00} = mn c_\LL^2
(\partial_3  \phi_3)^2
\;.
\end{equation}
Thus
\begin{equation}
\begin{split}
\label{eq:psi3result}
\psi_3 ={}&
\frac{C}{mn c_\LL^2}\,   
\int_{-\infty}^{\infty}\!d x^3\
T^{00}
\;,
\end{split}
\end{equation}
where
\begin{equation}
\label{eq:Cval}
C = \frac{9}{8}
\;.
\end{equation}
%

\section{Result of Ref.~\cite{EKN}}

The result (\ref{eq:psi3result}) is almost equivalent to the result of Ref.~\cite{EKN}, but the coefficient $C$ given Eq.~(\ref{eq:Cval}) in is different.  Ref.~\cite{EKN} has
\begin{equation}
\label{eq:CEKN}
C = \frac{d \log c_\LL}{d \log \rho}
\hskip 1 cm \textrm{Ref.~\cite{EKN}}
\;,
\end{equation}
where $\rho$ is the mass density of the material. The internal energy function $\widetilde U$ used in Ref.~\cite{EKN} includes any $\widetilde U$ consistent with the material being homogeneous and isotropic, so it includes the Hooke's Law function (\ref{eq:HookesLaw}) that I have used. I have analyzed sound waves in the material in its ground state, with density $\rho = mn$. In order to evaluate the derivative of $c_\LL$ with respect to $\rho$, we need to consider a material with a different undisturbed density, $\rho = mn b_0$ with $b_0 \ne 1$. This is simple. Instead of taking $R_a(x) = x_a + \varphi(x)$, where $\varphi(x)$ is a small perturbation, we now take $R_a(x) = b_0^{1/3} x_a + \widetilde\varphi(x)$ where $\widetilde\varphi(x)$ is a small perturbation. With this change, with $\widetilde\varphi(x) = 0$ there is a constant pressure in the solid:
\begin{equation}
P(b_0) = \left(\frac{3\lambda}{2}+\mu\right)b_0^{5/3}(b_0^{2/3} - 1)
\;.
\end{equation}
Now the speeds of longitudinal and transverse sound waves are
\begin{equation}
\begin{split}
c_\LL(b_0)^2 ={}& 
\frac{\lambda + 2\mu}{mn}\, b_0^{4/3} + \frac{3P(b_0)}{mn}\, b_0^{-1} 
\;,
\\
c_\LT(b_0)^2 ={}& 
\frac{\mu}{mn}\, b_0^{4/3} + \frac{P(b_0)}{mn}\, b_0^{-1}
\;.
\end{split}
\end{equation}
We can use this result to find the derivative of $c_\LL$ with respect to the density $mn b_0$ at $b_0 = 1$:
\begin{equation}
\begin{split}
\frac{d\log c_\LL }{d\log \rho}  ={}& 
\frac{13\lambda + 14\mu}{6(\lambda + 2 \mu)}
\;.
\end{split}
\end{equation}
This is the constant $C$ in Eq.~(\ref{eq:CEKN}) according to Ref.~\cite{EKN}. For arbitrary $\mu/\lambda$, it does not agree with the result of this paper, Eq.~(\ref{eq:Cval}).

In my judgement, the qualitative result in Eq.~(\ref{eq:psi3result}) is of some interest since it seems surprising, but the coefficient $C$ is not so important. Nevertheless, it may be if interest to understand how the difference could arise.

Ref.~\cite{EKN} allows the displacement field to depend on $x$ and $y$ as well as $z$ and analyzes the function $\partial_j\psi_j(\vec x, t)$. Using the momentum conservation equations, as in Eq.~(\ref{eq:psi3PDE}), one obtains for $\partial_j\psi_j$
\begin{equation}
\begin{split}
\label{eq:EKNeqn}
mn\left[\partial_0\partial_0 
- c_\LL^2 \partial_k\partial_k\right]
(\partial_j \psi_j) ={}& 
\partial_0\partial_j T_{\phi^2}^{j0}
+\partial_k\partial_j T_{\phi^2}^{jk}
\;,
\end{split}
\end{equation}
where $T_{\phi^2}^{j0}$ is the part of $T^{j0}$ in Eq.~(\ref{eq:momentumdensity}) that is quadratic in $\phi$ and $T_{\phi^2}^{jk}$ is the part of $T^{jk}$ in Eq.~(\ref{eq:Tjk}) that is quadratic in $\phi$. 

If we were to follow the analysis of Ref.~\cite{EKN}, we would consider a time average of $\psi(\vec x, t)$, defined by
\begin{equation}
\psi_\mathrm{av}(\vec x, t) = \int dt'\ h(t-t')\, \psi(\vec x, t')
\;,
\end{equation}
where $h$ is a function with $\int\! d\tau\,h(\tau) = 1$. We average over a long time: if $\omega$ is a typical frequency of the waves in the wave packet, then we ask that $\omega T \gg 1$, where $T = \int d\tau\, |\tau|\, h(\tau)$. Averaging gives us
\begin{equation}
\begin{split}
\label{eq:EKNeqn2}
mn\left(\partial_0\partial_0 
- c_\LL^2 \partial_k\partial_k\right)&
[\partial_j \psi_j]_\textrm{av} 
\\={}& 
\partial_0 \left[\partial_jT_{\phi^2}^{j0}\right]_\textrm{av}
+ \partial_k \left[\partial_jT_{\phi^2}^{jk}\right]_\textrm{av}
\;.
\end{split}
\end{equation}
With this averaging, $\partial_0[\partial_j\psi_j]_\mathrm{av}$ and $\partial_0 [\partial_jT_{\phi^2}^{j0}]_\textrm{av}$ are small, of order $1/(\omega T)$ times the same quantities before averaging. For this reason, if we followed Ref.~\cite{EKN}, we would drop the the time derivative terms in Eq.~(\ref{eq:EKNeqn2}). We would then have a differential equation involving only space derivatives. 

However, the averaging procedure smooths the oscillations in space similarly to how it smooths the oscillations in time, so that $\partial_k [\partial_j \psi_j]_\mathrm{av}$  and $\partial_k [\partial_jT_{\phi^2}^{j0}]_\textrm{av}$ are also small. Thus I am not able to justify this method of attack. The analysis given in the present paper has not neglected any terms in the equations of motion.

\section{Average velocity inside the wave packet}

Return now to Eq.~(\ref{eq:psi3result}). The wave packet takes a time $\Delta t$ to move past a given point. During that time, the matter moves a distance $-\psi_3$. Thus the average velocity of the matter was
\begin{equation}
\langle v \rangle = - \frac{\psi_3}{\Delta t}
\;.
\end{equation}
The length of the wave packet is $L = c_\LL \Delta t$. Thus
\begin{equation}
\langle v \rangle = - \frac{c_\LL \psi_3}{L}
\;.
\end{equation}
Using eq.~(\ref{eq:psi3result}), this is
\begin{equation}
\begin{split}
\label{eq:averagev}
\langle v \rangle ={}&
-\frac{9}{8 mn c_\LL}\,  
\langle T^{00} \rangle
\;,
\end{split}
\end{equation}
where 
\begin{equation}
\begin{split}
\label{eq:averageT00}
\langle T^{00} \rangle ={}&
\frac{1}{L} 
\int_{-\infty}^{\infty}\!d x^3\
T^{00}
\end{split}
\end{equation}
is the average energy density inside the wave packet.

To get some idea of what this velocity is in a simple example, we can use $\langle T^{00} \rangle =  \langle T^{03} \rangle / c_\LL$ and set
\begin{equation}
\left\langle
T^{03}
\right\rangle = 100\ \mathrm{W}/\mathrm{m}^2
\;.
\end{equation}
We can use material properties appropriate to a gel made mostly of water:
\begin{equation}
\begin{split}
mn ={}& 10^3\ \mathrm{kg}/\mathrm{m}^3
\;,
\\
c_\LL ={}& 1.5\times 10^3\ \mathrm{m}/\mathrm{s}
\;.
\end{split}
\end{equation}
For this example, we have
\begin{equation}
\langle v \rangle =
- 5\times 10^{-8}\ \mathrm{m}/\mathrm{s}
\;.
\end{equation}
Thus the atoms in our material move very slowly inside the sound wave packet.

\section{Mass deficit inside the wave packet}

We can relate $\psi_3$ directly to a mass deficit within the wave packet. We note that the mass density in the material is
\begin{equation}
mn \frac{\partial R_3}{\partial x^3}
= mn\left[1 + \partial_3 \phi_3 + \partial_3 \psi_3
\right]
\;,
\end{equation}
where we have included the terms proportional to $f^0$, $f^1$ and $f^2$ but omit higher order terms. The first term is the mass density of the undisturbed material. The remainder is then the mass density associated with the wave,
\begin{equation}
\rho_\LM
= mn\,\partial_3 \phi_3 + mn\, \partial_3 \psi_3 
\;.
\end{equation}
Then the mass per unit area associated with the wave is
\begin{equation}
\begin{split}
M ={}& mn \int_{c_\LL t - R}^{c_\LL t + R} dx^3\ 
\frac{\partial \phi_3}{\partial x^3}
+ mn \int_{c_\LL t - R}^{c_\LL t + R} dx^3\ 
\frac{\partial \psi_3}{\partial x^3}
\\
={}& 0
- mn\, \psi_3(t,c_\LL t - R)
\\
={}&
- \frac{9}{8 c_\LL^2}\,   
\int_{-\infty}^{\infty}\!d x^3\
T^{00}
\;.
\end{split}
\end{equation}
Here the first term in the first line vanishes because $\phi_3$ vanishes outside of the wave packet, but the second term is nonzero because $\psi_3$ does not vanish after the wave packet has passed but rather is given by Eq.~(\ref{eq:psi3result}). Thus the mass per unit area associated with the wave packet, $M$, is related to the energy per unit area carried by the wave, $E$, by
\begin{equation}
\label{eq:massresult}
M = - \frac{9}{8  c_\LL^2}\, E
\;.
\end{equation}
Except for the coefficient, this is the result of Ref.~\cite{EKN}.

\section{Connection to gravity}

Ref.~\cite{EKN}, titled ``Gravitational Mass Carried by Sound Waves,'' states that sound waves are affected by gravity and source gravity. In order to derive the connection to gravity, we could use the Lorentz invariant Lagrangian for an isotropic Hooke's Law material \cite{CFT}. Then we could generalize that to the corresponding Lagrangian that is invariant under general coordinate transformations \cite{CFT}. This formulation includes gravity through the coupling of the fields $R_a(x)$ to the metric tensor $g_{\mu\nu}(x)$. Then the mass that is transported by the sound wave interacts with gravity. However, in my opinion, using this much theory would only obscure the simple result that we see already in the nonrelativistic theory.

\bigskip
\acknowledgments{ 
This work was supported in part by the United States Department of Energy under grant DE-SC0011640. I thank A.~Esposito for helpful correspondence and T.~Cohen for helpful conversations.
}



\end{document}